\documentclass[onecolumn,showpacs,preprintnumbers,amsmath,amssymb]{revtex4}

\usepackage{epsf}
\usepackage{graphicx}  
\usepackage{dcolumn}   
\usepackage{bm}        

\newcommand{\be}{\begin{equation}}
\newcommand{\en}{\end{equation}}
\newcommand{\bea}{\begin{eqnarray}}
\newcommand{\ena}{\end{eqnarray}}

\begin{document}


\title{Braneworld cosmology in the sourced-Taub background}

 \author{Hongsheng Zhang\footnote{Electronic address: hongsheng@kasi.re.kr} }
  \affiliation{\footnotesize
 Korea Astronomy and Space Science Institute,  Daejeon 305-348, Korea }
 \affiliation{\footnotesize Department of Astronomy, Beijing Normal University,
Beijing 100875, China}
  \author{Hyerim Noh\footnote{Electronic address: hr@kasi.re.kr} }
 \affiliation{\footnotesize
 Korea Astronomy and Space Science Institute,
  Daejeon 305-348, Korea }

\begin{abstract}
 A new braneworld in the sourced-Taub background is proposed. The
 gravity field equations in the internal source region and external vacuum region
  are investigated, respectively. We find that the equation of state for the effective dark energy of a dust brane
  in the source region can cross the
 phantom divide $w=-1$. Furthermore, there is a drop on $H(z)$ diagram, which
 presents a possible mechanism for the recent direct data of $H(z)$.

\end{abstract}


\maketitle

\section{ Introduction}
    The brane world scenario
 is  an  impressive progress in high energy
 physics and cosmology in recent years. The basic idea of this scenario is that the standard model particles are
 confined to the 3-brane, while the gravitation can propagate in
 the whole spacetime \cite{brane}. Cosmology in the brane world scenario
 has been
 widely investigated. In the
 braneworld cosmology the bulk spacetime is assumed to be
 Schwarzschild-Anti-de Sitter (AdS)\cite{langlois} or just Minkowski \cite{dgpcosmology}.

 However, such bulks are not necessary for cosmology. In principle, we only require that the bulk space admits a 3-dimensional
 maximally symmetric spacelike submanifold, which serves as our
 space. Besides Schwarzschild-AdS (dS) and Minkowski, a 5-dimensional
 vacuum Taub space and the source of Taub space also admit 3-dimensional
 maximally symmetric spacelike submanifold, which  are proper backgrounds for brane cosmology.
 We will study the cosmology in the background of a sourced-Taub bulk.

 The source of Taub space is a long standing problem \cite{taub}. In a recent series
 of works \cite{self1, self2, self3}, we successfully find the
 source of the Taub space  both in 4-dimension and higher
 dimensions. The components of the source region are also
 preliminarily studied. These results strengthen the physical
 foundation of Taub space.

 In the next section we briefly review the former results of the
 source of the Taub solution. In section III, we shall explain our
 set up of the sourced Taub system with a moving brane. We study the
 cosmology of this set up in section IV. In section V we present our
 conclusion.

 \section{ The source of Taub space}
 A new class of static plane symmetric solution of Einstein field
 equation sourced by a perfect fluid was successfully found in 4-dimensional spacetime in \cite{self1}.
 This solution is identified as the source of the Taub space since there is a special family in this solution which can perfectly match
 to the Taub space.

 In \cite{self2}, we generalize the 4-dimensional source of Taub
 solution  to a higher dimensional one, which reads,
  \be
 ds^2=-e^{2az}dt^2+dz^2+e^{2[az+be^{az/(n-3)}]}d\Sigma^2,
  \label{metric}
 \en
 where
 \be
 d\Sigma^2=(dx^1)^2+(dx^2)^2+...+(dx^{n-2})^2.
 \en

 The above metric (\ref{metric}) is an exact solution of  Einstein field
equation sourced by a  perfect fluid,
 \be
 T=(\rho(z)+p(z))U\otimes U+p(z)g_n,
 \label{em}
 \en
 where $T$ denotes the energy momentum tensor of the fluid, $U$
 stands for 4-velocity of the fluid, $g_n$ denotes the n-dimensional metric
 tensor, and,
\be
  \rho=-\frac{a^2}{2}\frac{n-2}{(n-3)^2}\left[(n-3)^2(n-1)+2(n-2)^2be^{az/(n-3)}
  +(n-1)b^2e^{2az/(n-3)}\right],
  \label{rho}
 \en
 \be
  p=\frac{a^2}{2}\frac{n-2}{(n-3)}\left[(n-3)(n-1)+2(n-2)be^{az/(n-3)}
  +b^2e^{2az/(n-3)}\right].
  \label{p}
 \en
 In 4-dimensional case, the above energy momentum can be
 a phantom $\psi$ with dust and photon. The Lagrangian for the source is
 \be
 {\cal L}_{\rm source}=\frac{1}{2}
 \partial_{\mu}\psi\partial^{\mu}\psi-U(\psi)+{\cal L}_{\rm
 dust}+{\cal L}_{\rm photon},
 \en
 where the potential $U(\psi)$ is rather complicated, see
 \cite{self2} for details, ${\cal L}_{\rm
 dust}$ denotes the Lagrangian for the dust and ${\cal L}_{\rm
 photon}$ represents the Lagrangian of the photon. The substance of n-dimensional
 source is almost the same  as that of the 4-dimensional case. The reason of
 this similarity roots in the following fact: the only reasonable $U(\psi)$ is only a function of $z$ \cite{self2}.

 It is found that a family in the solution (\ref{metric}) can
 perfectly match to the n-dimensional Taub solution, and thus is the
 source of a n-dimensional Taub. We write the vacuum Taub metric as follows,
 \be
 ds^2=-z^{2\alpha}k^2dt^2+dz^2
 + z^{2\beta}l^2d\Sigma^2,
 \label{ntaub}
 \en
 where $\alpha=-\frac{n-3}{n-1},~\beta=\frac{2}{n-1}$. $k, l$ are two constants.
  If the vacuum region resides in $z>z_0$ and the source region inhabits
 $z<z_0$, the matching condition at the boundary $z=z_0$ requires,
 \bea
 k=\pm \frac{e^{az_0}}{z_0^{\alpha}},
 \label{azl}
 \ena
 \be
 l=\pm \frac{e^{az_0+be^{az_0/(n-3)}}}{z_0^{\beta}},
 \label{azb}
 \en
 and
  \be
    az_0=-\frac{n-3}{n-1}.
    \label{az0}
    \en
    For detailed discussions, see our original Letter
    \cite{self3}.

 \section{ The set up}
 We consider a 3-brane imbedded in a 5-dimensional bulk.
 The action includes the action of the bulk and the action of the brane,
 \be
 \label{action}
 S=S_{\rm bulk}+S_{\rm brane}.
 \en
 Here
 \be
 \label{actionbulk}
  S_{\rm bulk} =\int_{\cal M} d^5X \sqrt{-\det{(g_5)}}
\left( {1 \over 2\kappa^2 } R_5 +{\cal L}_{\rm bulk}  \right),
 \en
  where $X=(t,z,x^1,x^2,x^3)$ is the bulk coordinate, $x^1,x^2,x^3$ are the coordinates of the maximally symmetric space. ${\cal
  M}$, $\det{(g_5)}$, $\kappa$, $R_5$, ${\cal L}_{\rm bulk}$,
  denote the bulk manifold, the determinant of the
  bulk metric,  the 5-dimensional Newton constant,
   the 5-dimensional Ricci scalar, and the
  bulk matter Lagrangian, respectively. Note that the bulk matter Lagrangian ${\cal L}_{\rm
  bulk}$ can be a phantom with dust and photon.

 The action of the brane can be written as,
 \be
 \label{actionbrane}
 S_{\rm brane}=\int_{M} d^4 x\sqrt{-\det(g)} \left(
{\kappa^{-2}} K + L_{\rm brane} \right),
 \en
 where  $M$
  indicates the brane manifold, $\det(g)$ denotes the determinant of the
  brane metric, $L_{\rm brane}$ stands for the Lagrangian confined to the brane,
    and $K$ marks the trace of the second fundamental form of the
    brane. $x=(\tau, x^1,x^2,x^3)$ is the brane coordinate. Note
    that $\tau$ is not identified with $t$ if the the brane is not
    fixed at a position in the extra dimension $z=$constant. We will
    investigate the cosmology of a moving brane along the extra dimension
    $z$ in the bulk, and such that $\tau$ is different from $t$.

    We set the Lagrangian confined to the brane as follows,
     \be
    \label{labrane}
    L_{\rm brane}=-  \lambda + L_{\rm
m},
  \en
  where $\lambda$ is the brane tension and $L_{\rm m}$ denotes the
  ordinary matter, such as dust and radiation, located at the brane.
  We see that our set up is  a brane with tension and ordinary matter imbedded in
  a 5-dimensional vacuum bulk. Further we assume the bulk space is a sourced-Taub space. In this
  case, the n-dimensional metric in section II degenerates to
  ,
  \be
  g_5=e^{2f(z)}dt^2-dz^2-e^{2l(z)}\left[(dx^1)^2+(dx^2)^2+(dx^3)^2\right],
  \label{metric5}
  \en
  where
  \be
  f(z)=az,
  \en
  \be
  l(z)=az+be^{\frac{az}{2}},
  \en
  in the source region, and
  \be
  e^{2f(z)}=k^2z^{-1},
  \en
  \be
  e^{2l(z)}=m^2z,
  \en
  in the vacuum region, where $a,~b,~k,~m$ are constants.
  Correspondingly, the matching condition (\ref{azl},~\ref{azb},~\ref{az0})
  \be
   k=\pm \frac{e^{az_0}}{z_0^{\alpha}},
 \label{azl1}
  \en
  \be
 m=\pm \frac{e^{az_0+be^{az_0/2}}}{z_0^{\beta}},
 \label{azb1}
 \en
  and
  \be
  az_0=-\frac{1}{2},
  \label{az01}
  \en
  where the subscript $0$ denotes the value at the boundary of a
  quantity, and $\alpha=-1/2$, $\beta=1/2$.
  So there are only two free parameters, $a$ and $b$, in this set-up (source and external vacuum space), just as the
  same of the source. This is natural since the source should
  uniquely determine the external vacuum metric.
  Under this matching
  condition, the metric is only $C^1$ at the boundary.

  By considering a brane is moving in such a bulk, we investigate the cosmology of the brane in the next section.

  \section{cosmology}
   We take a method developed in \cite{bulkbase}. We suppose that a 3-brane is moving along $z$ direction, whose
  velocity is
  \be
  u^{\mu}=(u^0,u^1)=(\dot{t},\dot{z}),
  \en
   where the other 3 components of $u$ have been omitted, since they are
   just 0, and a dot stands for the derivative with respective to the proper time of the orbit of the brane in the bulk, $\tau$.
   The normalization condition of $u$ requires,
   \be
  g_5(u,u)=1,
  \label{unormal}
  \en
   which implies,
   \be
  e^{2f}\dot{t}^2-\dot{z}^2=1.
  \en
  The normal of the velocity satisfies,
 \be
 g_5(n,u)=0,
 \label{ortho}
  \en
 and
 \be
 g_5(n,n)=-1,
 \label{normal}
 \en
 which yield,
 \be
 e^{2f}n^0\dot{t}-n^1\dot{z}=0,
 \en
 and
 \be
 e^{2f}(n^0)^2-(n^1)^2=-1,
 \en
  respectively.
 Associating (\ref{ortho}), (\ref{normal}) and (\ref{unormal}), we
 obtain
 \be
 (n^1)^2=\dot{z}^2+1,
 \en
 \be
 (n^0)^2=e^{-4f}(\dot{z}^2+1)\frac{\dot{z}^2}{\dot{t}^2}.
 \en
  $n$ can be determined up to a sign, since direction of $n$ has not been uniquely selected.
  The induced metric on the brane reads,
  \be
  g=g_5+n\otimes
  n=d\tau^2-e^{2l(z)}\left[(dx^1)^2+(dx^2)^2+(dx^3)^2\right].
  \en
   The second form of the brane is the Lie derivation of the induced metric along the normal direction,
  \be
  K=\frac{1}{2}{\cal L}_{\vec{n}}g,
  \en
  whose spatial components read,
  \be
  K_{ij}=\frac{dl}{dz}n^1g_{ij},
  \en
  where $i,~j$ denote the spatial index of the brane universe.
 By imposing a $Z_2$ symmetry on the two sides of the brane, the matching condition across the brane yields,
  \be
  K_{\mu\nu}-Kg_{\mu\nu}=-\frac{\kappa^2}{2} s_{\mu\nu},
  \label{matching}
  \en
  where lower case of Greeks denotes the index of the quantity of the
  brane, which runs from 0 to 3, and $s_{\mu\nu}$ represents the energy momentum
  confined to the brane, which is defined as,
  \be
  s_{\mu\nu}\triangleq \frac{2}{\sqrt{-\det{(g)}}}\frac{\delta({\sqrt{-\det{(g)}} L_{\rm
  brane}})}{\delta g_{\mu\nu}},
  \label{embrane}
  \en
  where $L_{\rm
  brane}$ is given by (\ref{labrane}). The spatial components of
  (\ref{matching}) imply the Friedmann equation on the brane,
  \be
  \left(\frac{dl}{dz}\right)^2(1+\dot{z}^2)=\frac{\kappa^4}{36}
  \rho_{br}^2,
  \label{fried}
  \en
  where $\rho_{br}$ is the density of the brane, which is defined
  as,
  \be
  \rho_{br}\triangleq s_0^{0}.
  \label{rhobr}
  \en
  Here, the final form of the Friedmann equation is independent of
  the form of $f(z)$.
  We note that $\rho_{br}$ carries the effect of all the matters confined to the
  brane, including the vacuum energy $\lambda$. Only the Friedmann
  equation (\ref{fried}) is not enough to determine the evolution of
  the brane universe. The other essential equation is the Bianchi
  identity (continuity equation),
  \be
  \dot{\rho_{br}}+3H(\rho_{br}+p_{br})=0,
  \label{bianchi}
   \en
  where $H$ is the Hubble parameter, which is defined as
  \be
  H\triangleq \frac{\dot{\sqrt{z}}}{\sqrt{z}}.
  \label{hubble}
  \en
  $p_{br}$ denotes the pressure of the brane for a comoving
  observer,
  \be
  p_{br}\triangleq s_1^{1}=s_2^{2}=s_3^{3}.
  \en
  The continuity equation (\ref{bianchi}) also can be derived from the time
  component of the matching condition (\ref{matching}) \cite{langlois}.
  From (\ref{hubble}) we see  that the spatial coordinate $\sqrt{z}$ plays
  the role of the scale factor of the brane universe. The Friedmann
  equation (\ref{fried}) is a special case of the most general form
  of brane world model \cite{cov}. In this article, we first study a brane imbedded in the
   sourced Taub space.

  In the source region of the bulk, $l(z)=az+be^{\frac{az}{2}}$,
  hence the Friedmann equation (\ref{fried}) becomes,
  \be
  \frac{1}{4z^2}+H^2=\frac{\kappa^4\rho_{br}^2}{144a^2z^2(1+\frac{b}{2}e^{\frac{az}{2}})^2}.
  \label{sfried}
  \en
 In the external vacuum region, $e^{2l(z)}=m^2z$, the Friedmann equation (\ref{fried}) becomes,
 \be
  \frac{1}{4z^2}+H^2=\frac{\kappa^4}{576}\rho_{br}^2.
  \label{vfried}
  \en
 Two comments on the Friedmann equation are listed as follows:
 \begin{itemize}
  \item [1.]
   A term $1/z^2$, which is very alike the radiation term, but it is not the
   radiation, appears. This term is often called dark radiation in the
   literatures, which is the contraction of bulk Weyl tensor. In a spherically symmetric bulk,
   the physical sense of this term is the gravitational mass of the
   bulk space. In the present set up, it is a free term since $z$ is
   free adjusting coordinate. This confirms the fact that the Taub
   space has no proper definition of gravitational mass since it is
   not asymptotically flat.
   \item [2.]
   The present model only admits a spatially flat universe since
   Taub  space does not admit other type of splittings with a 3
   dimensional space-like submanifold. It is clear that the
   5-dimensional sourced-Taub (\ref{metric5}) admits 6 space-like
   Killing vectors, including 3 translational vectors and 3
   rotational ones, which span a 6 dimensional Euclidean group
   $E_6$.
     \end{itemize}
   We expect the present braneworld model can say something on the dark energy problem.
   First, we try to obtain the condition for cosmic acceleration in this model.
    Since the Friedmann equation seems fairly complicated, we deduce
   the acceleration equation by a very general function of the density of $\rho_{br}$ and scale factor
   $R\triangleq\sqrt{z}$,
   \be
   H^2=\phi (\rho_{br},R).
   \en
  Associating with continuity equation (\ref{bianchi}), one derives
  \be
   \frac{2\ddot{R}}{{R}}=-3(\rho_{br}+p_{br})\frac{\partial \phi}{\partial
   \rho_{br}}+R\frac{\partial \phi}{\partial
   R}+2\phi.
   \en
 One sees that the acceleration of the universe is only determined
 by the density and pressure iff
 \be
  \phi(\rho_{br},R)=\psi(\rho_{br})+\frac{C}{R^2},
  \label{phi}
  \en
  where $C$ is a constant, otherwise the scale factor $R$ will explicitly appear in the representation of
  the acceleration. It is evident that $\frac{C}{R^2}$ is just the
  spatial curvature term.
  Both the Firedmann equations in the source region (\ref{sfried})
  and in the vacuum region (\ref{vfried}) can not be written in the form of (\ref{phi}),
  thus the acceleration has to be an explicit function of $R$. In
  the source region, the acceleration becomes
  \be
  \frac{2\ddot{R}}{{R}}=-6(1+\frac{p_{br}}{\rho_{br}})(H^2+\frac{1}{4R^4})+2H^2+R\frac{\partial \phi}{\partial
   R},
  \en
  where
  \be
  R\frac{\partial \phi}{\partial
   R}=\frac{1}{R^4}\left\{1-\frac{\kappa^4\rho_{br}^2}{144a^2(1+\frac{b}{2}e^{aR^2/2})^3}\left[4+b(2+aR^2)e^{aR^2/2}\right]\right\}.
   \en
  In the vacuum region, the acceleration becomes
  \be
  \frac{\ddot{R}}{{R}}=-\frac{\kappa^4}{144a^2}(2\rho_{br}+3p_{br})\rho_{br}+\frac{1}{4R^4}.
  \label{friedv}
  \en
  At the late-time, the dark radiation term is reasonably committed. So, if the brane is moving in the vacuum region of
  the bulk, the acceleration condition becomes
  \be
   2\rho_{br}+3p_{br}<0,
   \en
   if we require $\rho_{br}>0$ (However, for an asymptotic anti-de Sitter brane, it is incorrect since the total density may be negative
   in the late time universe. Here we do not consider this case.). It is a stronger condition than
   that of the standard case $\rho+3p<0$ in the sense that it needs a
   smaller pressure for the same density.

    By using (\ref{labrane})
   and (\ref{rhobr}), we obtain
   \be
     \rho_{br}=\rho_m+\lambda,
      \label{decorho}
     \en
     where $\rho_m$ is the density of matter confined to the brane. In
     the following text, we consider the case $\rho_m\sim 1/R^3$,
     which  means that we consider a dust brane with tension, but without exotic
     matter.
     Substituting (\ref{decorho}) into (\ref{sfried}), we reach
     \be
     H^2=-\frac{1}{R^4}+\frac{\kappa^4\rho_m^2}{144a^2R^4(1+\frac{b}{2}e^{aR^2/2})^2}+
     \frac{\lambda
     \kappa^4}{72a^2}\frac{\rho_m}{R^4(1+\frac{b}{2}e^{aR^2/2})^2}+
     \frac{\lambda^2\kappa^4}{144a^2R^4(1+\frac{b}{2}e^{aR^2/2})^2}.
     \label{expfriedin}
     \en

    However, the Friedmann equation should recovers to the standard one at the time when CMB decoupling and structure
   formation occur. (\ref{expfriedin}) cannot recover to the standard one at the early time,
   so it is proper to consider a brane moving in a vacuum bulk in
   the early time, where the tension of the brane is required. In
   the vacuum region, by using (\ref{decorho}) the Friedmann equation
   (\ref{vfried}) becomes,
   \be
   H^2=-\frac{1}{R^4}+\frac{\kappa^4}{576}\left(\rho_m^2+2\rho_m\lambda+\lambda^2\right).
   \en
   Then, as usual in the brane world model, we define,
   \be
   \frac{8\pi G_{eff}}{3}=\frac{\kappa^4\lambda}{288}.
   \label{geff2}
   \en
   Comparing with the previous brane models, we do not include a bulk
  cosmological constant since we do not find a proper source of
  Taub-(A)dS space yet.
   This definition of $G_{eff}$ is the same as the former definition in
   brane world model up to a factor \cite{langlois}. It has been
   investigated in detail except a dark radiation term, which is
   unimportant in the late time universe.

   Our scenario is that the brane moves in a vacuum bulk in the high
   redshift region and enters in  the source region in some low
   redshift region. And the accelerating universe is driven by the
   source matter in the bulk. For recovering the standard cosmology
   in the middle redshift region and high redshift region, we need
   the brane tension.

   For convenience of numerical calculation, we write the Firedmann equation in the form of
   evolution with respect to the redshift $\xi$. In the source
   region,
   \be
   \frac{H^2}{H_p^2}=-\Omega_{dr}(1+\xi)^4+\frac{1}{2\Omega_{\lambda}} \left[\Omega_{mp}(1+\xi)^3+\Omega_{\lambda}\right]^2,
   \label{h2hp}
   \en
   where
   \be
   \Omega_{dr}=\frac{1}{4H_p^2R_p^4},
   \en
   \be
   \Omega_{mp}=\frac{8\pi B\rho_p}{3H_p^2},
   \en
   and
   \be
   \Omega_{\lambda}=\frac{8\pi B\lambda}{3H_p^2},
   \en
 where the subscript $p$ labels the present epoch of a physical
  quantity, $B$ is defined as
  \be
  \frac{8\pi B}{3}=\frac{\rho_m
     \kappa^4}{144a^2R^4(1+\frac{b}{2}e^{aR^2/2})^2}.
     \label{f}
  \en
   We should not confuse the symbol of the present value $p$ with the value of the quantity
  at the boundary, for which we use $0$ to denote.

   In the vacuum region, the Friedmann equation keeps the same
   form. But $B$ is replaced by $G_{eff}$ in(\ref{geff2}). In contrast to
   the previous brane models, we find an interesting property of this
   brane world model: The equation of state (EOS) of the effective dark
   energy (defined as the ratio of
 pressure to energy density) can cross the phantom divide $w=-1$ on a dust-brane with tension.

  The crossing $-1$ behavior of EOS is a deep problem and a big challenge to the fundamental physics,
   which is aroused by more accurate data:
  the recent analysis of the type Ia supernovae data
  indicates that the time varying dark energy gives a better
  fit  than a cosmological constant, and in particular, the (EOS)
     $w$  may cross $-1$ \cite{vari}.
 The dark energy with $w<-1$ is called phantom dark
 energy~\cite{call}, for which all energy conditions are
 violated. Here, it should be noted that the possibility that the dark energy behaves as
  phantom today is yet a matter in debate: the
 observational data, mainly those coming from the type Ia supernovae
 of high redshift and cosmic microwave background, may lead to different conclusions
 depending on what
 samples are selected, and what statistical analysis is
 applied \cite{jassal}. By contrast, other researches imply that all
 classes of dark energy models are comfortably allowed  by those
 observations \cite{contra}. Presently all observations
 seem  not to rule out the possibility of the existence of matter with $w
 <-1$. Even in a model in which the Newton constant is evolving with
 respect to the redshift $z$, the best fit w(z) crosses the phantom divide $w=-1
 $ \cite{Nesseris}. Hence the phenomenological model for phantom dark energy
 should be considered seriously.
  To obtain $w <-1$, scalar field with a negative kinetic term,
  may be a simplest realization. The model with phantom matter has been investigated
  extensively \cite{phantom}, and a test of such matters in solar system, see \cite{solar}.  However,
 the EOS of phantom  field is  always
 less than $-1$  and can not cross $-1$.  It is easy to understand that if we put
 two
 scalar fields into the model, one is an ordinary scalar and the other
 is a phantom: they dominate the universe by turns, under this
 situation the effective EOS can cross $-1$ \cite{quintom}.   It is worthy to point out that
  there  exist some interacting models, in which the effective EOS of dark energy
  crosses $-1$ \cite{interact}.
   More recently
  it has been found that crossing $-1$ within one scalar field
  model is possible, the cost is that the action contains higher derivative
  terms~\cite{Li} (see also \cite{Binflation}). Also it is found
  that such a crossing can be realized
   without introducing ordinary scalar or phantom component in a
 Gauss-Bonnet brane world with induced gravity, where a four
 dimensional curvature scalar on the brane and a five dimensional
 Gauss-Bonnet term in the bulk are present \cite{hs1}.

 To explain the observed evolving EOS of effective dark energy,
  we calculate the equation of state $w$ for the effective
  ``dark energy" caused by the tension and term representing
  brane world effect
  by comparing the modified Friedmann equation in
  the brane world scenario and the standard Friedmann equation in general
  relativity, because all observed features of dark energy are
  ``derived" in general relativity.
   Note that the standard Friedmann equation in a
 four dimensional spatially flat universe can be written as
 \be
 H^2=\frac{8\pi G}{3} (\rho_{m}+\rho_{de}),
 \label{genericF}
 \en
 where the first term in RHS of the above equation represents the dust matter and the second
 term stands for the effective dark energy, and $G$ is the Newton constant. Comparing (\ref{genericF})
 with (\ref{fried}), one obtains the density of effective dark
 energy,
 \be
 \rho_{de}=-\frac{1}{4R^4}+\frac{\kappa^4\rho_{br}^2}{144a^2R^4(1+\frac{b}{2}e^{\frac{aR^2}{2}})^2}-\frac{8\pi G}{3} \rho_{m}.
 \label{rhode}
 \en

 Since the dust matter obeys the continuity equation
 and the Bianchi identity keeps valid, dark energy itself satisfies
  the continuity equation
 \be
 \frac{d\rho_{de}}{dt}+3H(\rho_{de}+p_{eff})=0,
 \label{em}
 \en
 where $p_{eff}$ denotes the effective pressure of the dark energy.
 And then we can express the equation of state for the dark
 energy as
   \be
  w_{de}=\frac{p_{eff}}{\rho_{de}}=-1+\frac{1}{3}\frac{d \ln \rho_{de}}{d \ln
  (1+\xi)}.
   \en
  Clearly, if $\frac{d \ln \rho_{de}}{d \ln
  (1+\xi)}$ is greater than 0, dark energy evolves as quintessence; if $\frac{d \ln \rho_{de}}{d \ln
  (1+\xi)}$ is less than 0, it evolves as phantom; if $\frac{d \ln \rho_{de}}{d \ln
  (1+\xi)}$ equals 0, it is just cosmological constant. In a more
  intuitive way, if $\rho_{de}$ decreases and then increases
  with respect to redshift (or time), or increases and then
  decreases, which implies that EOS of dark energy crosses phantom divide.
  The more important reason why we use the density to describe
  property of dark energy is that the density is
 more closely related to observables, hence is more tightly
 constrained for the same number of redshift bins used \cite{wangyun}.

  We find a concrete example in which the EOS of the effective dark
  energy crosses $-1$. For the numerical improvement, we should nail
  down the parameters in this model, which satisfy the requirements of theory and observation.
 For the theoretical aspect, the boundary condition of the source
 region yields \cite{self3},
 \be
 b=-4e^{-\frac{az_0}{2}},
 \label{b}
  \en
 and (\ref{az01}). We only consider the large branch in \cite{self3},
 hence $b$ is fixed by (\ref{az01}) and (\ref{b}). Therefore, to this
 brane model imbedded in sourced Taub background itself, we have
 only one free parameter. It is $a$ or $z_0$, which implies the
 thickness of the source region. The partition of dark radiation should be smaller than
 that of cosmic microwave background (CMB), otherwise the universe will bounce at some high redshift, thus
 $\Omega_{dr}<10^{-4}$. Here we set $\Omega_{dr}=10^{-5}$, $R_p=1$,
 $\Omega_{mp}=0.27$ and $\Omega_\lambda=1.408$. The present part of baryon density is
 approximately $4\%$ \cite{wmap}, and we have several evidences of the existence
 of the non-baryon dark matter, which are independent of $\Lambda$CDM model
 \cite{darkmatter}. Here we take $\Omega_{mp}$ as the same value of
 the best fit of $\Lambda$CDM model. $\Omega_{\lambda}$ is derived
  from (\ref{h2hp}) at $\xi=0$. Figure 1 illustrates the evolution of
 the effective density. It is clear that the EOS of the dark energy
 crosses $-1$ at $\xi=0.2\sim 0.3$. This point is also confirmed by the direct plot
 of the EOS of effective dark energy in figure 2, which well explains the
 observations \cite{vari}.

  \begin{figure}
\centering
 \includegraphics[totalheight=2.3in, angle=0]{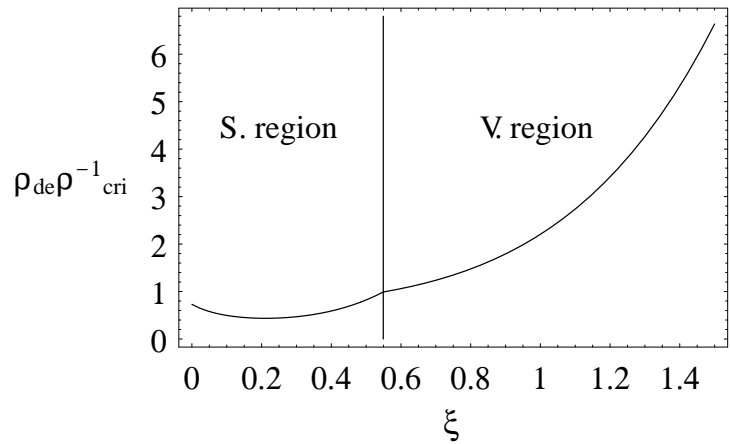}
\caption{The evolution of the density for the effective dark energy
 $\rho_{de}$ with respect to the redshift $\xi$, where $\rho_{\rm cri}$ denotes the
 present critical density. The left part (low redshift region)
is the source region, and the right part (high redshift region) is
the vacuum region. In this figure $a=-1.2$. The boundary of the
source region inhabits at $R_0=0.645$. The brane sweeps across the
boundary when $\xi=0.549$.}
 \label{rhode}
 \end{figure}

 \begin{figure}
\centering
 \includegraphics[totalheight=2.3in, angle=0]{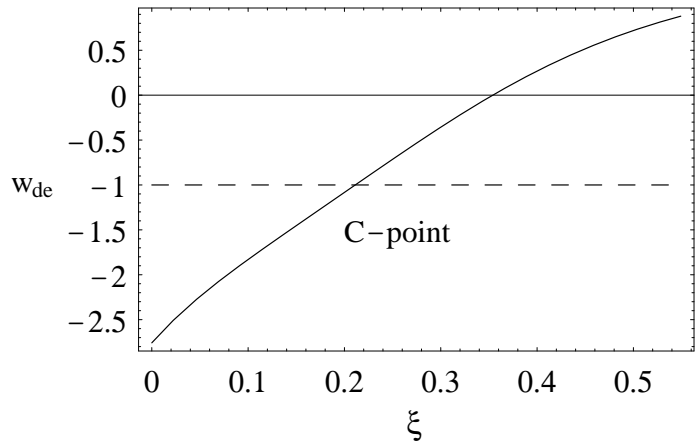}
\caption{The evolution of $w_{de}$ with respect to $\xi$. The point
 of crossing $-1$ (C-point) appears at $\xi= 0.2\sim 0.3$.
  The parameters
are the same as Figure 1.}
 \label{rhode}
 \end{figure}

 The most significant parameters from the viewpoint of
  observations is the Hubble parameter $H(z)$, which carries the total
  effects of cosmic fluids. Except the indirect data of $H(z)$, such as the luminosity distances of supernovae,
  the direct $H(z)$ data appear in recent years, which can be used to
  explore the fine structures of the Hubble expansion history \cite{hz}. There is an important new feature of $H(z)$ data which
  is not implied by the previous indirect observations of Hubble parameter: It decreases
   with respect to the redshift $\xi$ at redshift $\xi\sim 0.15$
   and $\xi\sim 1.5$, which means that the total
   fluid in the universe behaves as phantom. We find that the present
   brane model also can partly realize this property with the same parameters of the example in which the EOS of the
   effective dark energy crosses $-1$. Figure 3 illustrates
   the evolution of the Hubble parameter. At $\xi\sim 0.15$, there is
   a clear drop on $H(z)$ diagram in Figure 3.

  \begin{figure}
\centering
 \includegraphics[totalheight=2.5in, angle=0]{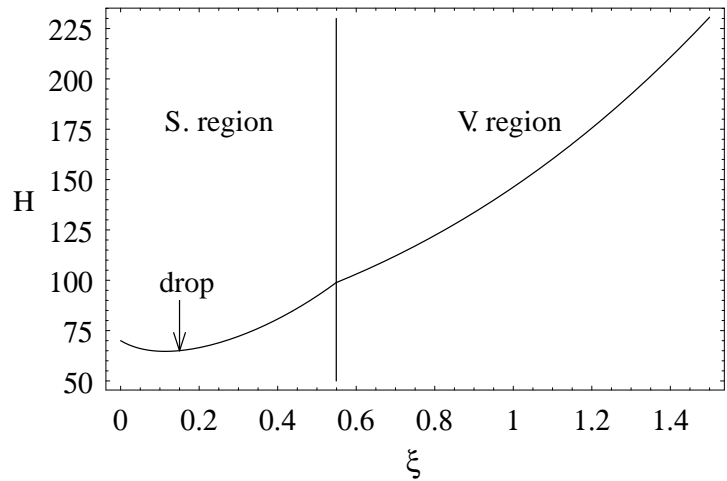}
\caption{The evolution of the Hubble parameter with respect to
redshift. The unit of $H$ is $km~s^{-1}\,Mpc^{-1}$. The parameters
are the same as Figure 1.}
 \label{rhode}
 \end{figure}

 Figure 1 and Figure 3 illuminate that there are cusps at the
 boundary of vacuum and source region, which is related to the fact
 that the metric on the boundary is only $C^1$, while second derivatives with respective to coordinates are involved in  Einstein
 equation.

 \section{conclusion}

 We investigate the brane world model in the
 sourced-Taub background. The Friedmann equations are obtained both
 in the source region and  the vacuum region.

 The recent observations imply that the EOS of dark energy may cross $-1$ and
 there may be some drops on the Hubble diagram.  These two phenomena are serious challenges
 to the
  physical cosmology. In the braneworld model with a sourced Taub background, for a reasonable parameter set
   the EOS of the effective dark
  energy can cross $-1$ in when the brane is moving in the source
  region. And furthermore, the Hubble parameter has a drop in the
  low redshift region (the source region).

  In the present work, we do not include a bulk cosmological
  constant and thus the brane tension can not be too large. When the brane is moving in the
  vacuum region, the quadratic term of  $\rho$ will dominate the linear term $\rho$ before
  long. In the high redshift region of the vacuum bulk, the model is
  treated as a toy model. Our main aim of this article is to study
  the behavior in the source region. The detailed evolution of the brane universe in sourced Taub background
  in the high redshift region needs to be investigated further in the
  future.

  {\bf Acknowledgments.}
 We are grateful to the anonymous referee for several valuable comments.
 H.Noh was supported by grant No. C00022 from the Korea Research
 Foundation.

\end{document}